\documentclass[preprint1]{aastex}
\usepackage{amsmath}
\slugcomment{}

\newcommand{\be}{\begin{equation}}
\newcommand{\ee}{\end{equation}}

\shorttitle{GRB Hubble Diagram}
\shortauthors{Wei, Wu \& Melia}

\begin{document}

\title{The Gamma-Ray Burst Hubble Diagram and its Implications for Cosmology}
%\author{Jun-Jie Wei\thanks{E-mail: jjwei@pmo.ac.cn} \hskip0.07in and Xue-Feng Wu\thanks{E-mail:
%xfwu@pmo.ac.cn}\\
%Purple Mountain Observatory, Chinese Academy of Sciences, Nanjing 210008, China}
%\author{and\\
%Fulvio Melia\thanks{John Woodruff Simpson Fellow. E-mail: melia@as.arizona.edu}\\
%Department of Physics, The Applied Math Program, and Department of Astronomy,\\
%The University of Arizona, AZ 85721, USA}
\author{Jun-Jie Wei\altaffilmark{1}, Xue-Feng Wu\altaffilmark{1,3,4}, and Fulvio Melia\altaffilmark{2}}
\altaffiltext{1}{Purple Mountain Observatory, Chinese Academy of Sciences, Nanjing 210008,
China; jjwei@pmo.ac.cn, xfwu@pmo.ac.cn.}
\altaffiltext{2}{Department of Physics, The Applied Math Program, and Department of Astronomy,
The University of Arizona, AZ 85721, USA; melia@as.arizona.edu.}
\altaffiltext{3}{Chinese Center for Antarctic Astronomy, Nanjing 210008, China.}
\altaffiltext{4}{Joint Center for Particle, Nuclear Physics and Cosmology, Nanjing
University-Purple Mountain Observatory, Nanjing 210008, China.}

\begin{abstract}
In this paper, we continue to build support for the proposal to use gamma-ray bursts
(GRBs) as standard candles in constructing the Hubble Diagram at redshifts beyond the
current reach of Type Ia supernova observations. We confirm that correlations
among certain spectral and lightcurve features can indeed be used as luminosity
indicators, and demonstrate from the most up-to-date GRB sample appropriate
for this work that the $\Lambda$CDM model optimized with these data is
characterized by parameter values consistent with those in the concordance
model. Specifically, we find that $(\Omega_m,\Omega_\Lambda)\approx
(0.25_{-0.06}^{+0.05}, 0.75_{-0.05}^{+0.06})$, which are consistent, to
within $1\sigma$, with $(0.29,0.71)$ obtained from the 9-yr WMAP data.
We also carry out a comparative analysis between $\Lambda$CDM and the
$R_{\rm h}=ct$ Universe and find that the optimal $\Lambda$CDM
model fits the GRB Hubble Diagram with a reduced $\chi^2_{\rm dof}\approx
2.26$, whereas the fit using $R_{\rm h}=ct$ results in a $\chi^2_{\rm dof}\approx
2.14$. In both cases, about $20\%$ of the events lie at least $2\sigma$
away from the best-fit curves, suggesting that either some contamination
by non-standard GRB luminosities is unavoidable, or that the errors and
intrinsic scatter associated with the data are being underestimated. With
these optimized fits, we use three statistical tools---the Akaike Information
Criterion (AIC), the Kullback Information Criterion (KIC), and the Bayes
Information Criterion (BIC)---to show that, based on the GRB Hubble Diagram,
the likelihood of $R_{\rm h}=ct$ being closer to the correct model is $\sim 85-96\%$,
compared to $\sim 4-15\%$ for $\Lambda$CDM.
\end{abstract}


\section{Introduction}
For a given class of sources whose luminosity is accurately known, one may
construct a Hubble Diagram (HD) from the measurement of their distance
versus redshift. Such a relationship can be a powerful tool for probing the
cosmological expansion of the Universe, but only if these sources truly
function as standard candles.
The cosmic evolution depends critically on its constituents, so measuring
distances over a broad range of redshifts can in principle place meaningful
constraints on the assumed cosmology. The discovery of dark energy was
made using this method, in which the sources---Type Ia supernovae---are
transient, though with a well-defined  luminosity versus color and light-curve
shape relationships (Riess et al. 1998; Perlmutter et al. 1998, 1999; Garnavich
et al. 1998; Schmidt et al. 1998). Of course, one must also assume that
the power of distant explosions can be standardized against those seen
at much lower redshifts.

The use of Type Ia SNe has been quite impressive, so one may wonder
why there would be a need to seek other kinds of standard candle. But
the reality is that several important limitations mitigate the overall impact
of supernova studies. For example, even excellent space-based platforms,
such as SNAP (Scholl et al. 2004), cannot observe these events at redshifts
$\ga 1.8$. And this is quite limiting because much of the most interesting
evolution of the Universe occurred well before this epoch. In addition,
the determination of the supernova luminosity cannot be carried out
independently of the assumed cosmology, so the Type Ia SN data
tend to be compliant to the adopted expansion scenario (Melia 2012a).
The fact that so-called ``nuisance" parameters associated with the
data need to be optimized along with the variables in the model
itself weakens any comparative analysis between competing
cosmologies. There is therefore much more to learn about the
Universe's history than one can infer from Type Ia SNe alone.

In recent years, several other classes of source have been proposed
as possible standard candles in their own right. Most recently, the
discovery that high-$z$ quasars appear to be accreting at close
to their Eddington limit (see, e.g., Willott et al. 2010), has made it possible
to begin using them to construct an HD at redshifts beyond $\sim 6$
(Melia 2012b). It has also been suggested that Gamma-ray Bursts
(GRBs) may be suitable for constructing an HD at intermediate redshifts,
$1\la z \la 6$, between the Type Ia SN and high-$z$ quasar regions.

The possible use of GRBs as standard candles started to become
reality after Norris et al. (2000) found a tight correlation
between the burst luminosity $L$ and the spectral lag $\tau_{\rm lag}$.
Some other GRB luminosity indicators have been widely discussed
in the literature. Amati et al. (2002) discovered a relationship between
the isotropic equivalent gamma-ray energy $(E_{\gamma,{\rm iso}})$ and the
burst frame peak energy in the GRB spectrum $(E_{\rm p})$, but the relation
may be the result of selection effects (see, e.g., Kocevski 2012; Collazzi et al.
2012). Similarly, the isotropic peak luminosity $(L_{\gamma,{\rm iso}})$
is also found to be correlated with $E_{\rm p}$ in the burst frame
(Schaefer 2003a; Wei \& Gao 2003; Yonetoku et al. 2004). Ghirlanda et al.
(2004a) replaced $E_{\gamma,{\rm iso}}$ with collimation-corrected gamma-ray
energy $(E_\gamma)$, and claimed a tighter correlation between $E_{\rm p}$
and $E_\gamma$. Since $E_\gamma=E_{\gamma,{\rm iso}}(1-\cos\theta)$, where
$\theta$ is the jet half-opening angle, one may reliably estimate the
isotropically equivalent energy $E_{\gamma,{\rm iso}}$ and use this to
infer a distance. Liang \& Zhang (2005) introduced the concept of optical
temporal break time $t_{\rm b}$, and discovered a strong dependence of
$E_{\gamma,{\rm iso}}$ on $E_{\rm p}$ and $t_{\rm b}$ without imposing
any theoretical models.

Earlier, Schaefer (2003b) had constructed the first GRB HD based
on nine events using two luminosity indicators, and this was
followed by Bloom et al. (2003a), who published a GRB HD with
16 bursts, assuming that the burst energy is a constant after
correcting for the beam angle. Some authors attempted to show
the HD for an observed GRB sample plotted against a theoretical HD
calculated using the same cosmological parameters (e.g., Dai et al.
2004; Liang \& Zhang 2005; Xu et al. 2005). These and other attempts
at constructing a GRB HD were made with only a small fraction of
the available data, using only one or two luminosity indicators.
Unfortunately, all of them had error bars that were too large
to provide useful constraints on cosmology. Schaefer (2007)
made use of five luminosity indicators and successfully
constructed a GRB HD with 69 events.

The feasibility of using this method became better grounded
when Dai et al. (2004) used the correlation found by Ghirlanda
et al. (2004a) to place tight constraints on the cosmological
parameters. Because of the current poor information
on low-$z$ GRBs, the Ghirlanda relation necessarily depends
on the assumed cosmology. Other authors attempted to circumvent
the circularity problem by using a less model-dependent approach (Ghirlanda
et al. 2004b, 2006; Firmani et al. 2005; Xu et al. 2005; Liang \&
Zhang 2005, 2006; Wang \& Dai 2006; Su et al. 2006; Li et al. 2008;
Qi et al. 2008a, 2008b; Liang et al. 2008; Wang et al. 2011). In the end,
however, as is also true for Type Ia SNe, the correlation must still be
recalibrated for each different model because the best-fit correlation
depends on the cosmology adopted to derive the burst luminosities.

Of course, even with the emergence of more precise luminosity
indicators, one must still deal with several significant challenges
when trying to use GRBs to construct an HD. The luminosity
of these bursts, calculated assuming isotropy, spans about
4 orders of magnitude (Frail et al. 2001). However, there is strong
observational evidence (e.g., the achromatic break in the
afterglow lightcurve) that the burst emission is collimated
into a jet with the aforementioned aperture angle $\theta$
(Levinson \& Eichler 1993; Rhoads 1997; Sari et al. 1999;
Fruchter et al. 1999). When one corrects for the collimation
factor $(1-\cos\theta)$, the gamma-ray energy tends to
cluster around $E_\gamma\sim 10^{51}$ ergs, but the
dispersion ($\sim 0.5$ dex) is still too large for these
measurements to be used for cosmological purposes.
For example, Wang et al. (2011) found that the updated
$\tau_{\rm lag} - L$ correlation has a large
intrinsic scatter.  And Lu et al. (2012) presented a time-resolved
$E_{\rm p}-L_{\gamma,{\rm iso}}$ correlation analysis and showed
that the scatter of the correlation is comparable to that of the
time-integrated relation. So in the end, theses luminosity correlations
may also not be suitable for cosmological purposes.
This is why much effort has been expended since 2004
in finding other indicators from the GRB spectrum that provide
more precise constraints on the luminosity.

One useful application of these ideas involves the use
of $E_{\rm p}$ and $t_{\rm b}$ (the so-called jet break
time) as a measure of $\theta$ to determine $E_{\gamma,{\rm iso}}$
(Liang \& Zhang 2005). In this paper, we will follow this
approach with two distinct goals in mind. First, several new
GRB events have been detected in recent years that have
spectral and lightcurve features (such as $t_{\rm b}$) with
sufficient quality to help improve the previously assembled
correlations. Second, and foremost, we wish to use this
relatively new probe of the Universe's expansion to directly
test the $R_{\rm h}=ct$ Universe (Melia 2007;
Melia \& Shevchuk 2012) against the data and to see
how its predictions compare with those of the $\Lambda$CDM
cosmology.

The $R_{\rm h}=ct$ Universe is a Friedmann-Robertson-Walker
cosmology that strictly adheres to the simultaneous requirements
of both the Cosmological principle and Weyl's postulate. Whereas
$\Lambda$CDM guesses  the constituents of the Universe and
their equation of state, and then predicts the expansion rate
as a function of time, $R_{\rm h}=ct$ acknowledges the fact
that no matter what these constituents are, the total energy
density in the Universe gives rise to a gravitational horizon
coincident with the better known Hubble radius. But because
this radius is therefore a proper distance, the application of
Weyl's postulate forces it to always equal $ct$. Thus, on
every time slice, the energy density must partition itself
among its various constituents in such a way as to always
adhere to this constraint, which also guarantees that the
expansion rate be constant in time.  As we shall see,
with all its complexity, $\Lambda$CDM actually mimics the
$R_{\rm h}=ct$ Universe when its free parameters are
optimized to produce a best fit to the cosmological data.

In the next section, we will describe the data we will use and
our method of analysis. We will then first assemble the GRB
HD in the context of the standard model, $\Lambda$CDM,
and demonstrate that the best-fit parameters obtained by
fitting its predicted luminosity distance to the GRB observations
very closely mirror those obtained through the analysis
of Type Ia SNe. In \S~4, we will introduce the $R_{\rm h}=ct$
Universe and provide a brief overview of its current status as
a viable cosmology. We will then construct the GRB HD for
this expansion scenario, which is much easier to do than
for $\Lambda$CDM because the former has only one free
parameter---the Hubble constant $H_0$. Finally, we will
directly compare the results of our fits to the data with
both $\Lambda$CDM and $R_{\rm h}=ct$.

\section{Observational data and Methodology\label{sec:intro}}
Our GRB sample includes 33 bursts with a measurement of the redshift \emph{z},
the spectral peak energy $E_{\rm p}$, and the jet break time $t_{\rm b}$ seen
in the optical afterglow. In assembling this sample, we required that the members
have an independent \emph{z}, that a spectral fit be available, and that a
jet-break characteristic be present in the optical band lightcurve. Note that in order to
preserve homogeneity, we did not include those bursts whose afterglow
break times were observed in the radio band (e.g., GRB 970508) or in the
X-ray band (e.g., GRBs 050318, 050505, 051022, 060124, 060210)
but were not seen in the optical band. We also excluded those bursts
whose \emph{z} or $E_{\rm p}$ were not directly measured. For example, because of
the narrowness of the \emph{Swift}/BAT band, the spectrum of GRB 050904 can be described
using a simple power law (Tagliaferri et al. 2005), so we do not know the real $E_{\rm p}$.
Some bursts with reported \emph{z} and $E_{\rm p}$ were also not included for a variety
of reasons: GRB 050820A does not have a well measured $t_{\rm b}$; there is only one
observed datum in the last decay phase of its optical lightcurve (see Fig. 4 of
Cenko et al. 2006). The optical lightcurve of GRB 060418 is characterized by an initial sharp
rise, peaking at 100 $-$ 200 s, with a subsequent power-law decay
(Molinari et al. 2007). GRB 060418 does not have a jet-break characteristic. Racusin
et al. (2008) found that the observed afterglow of GRB 080319B can be interpreted
using a two-component jet model, so this burst may not be a ``Gold" jet-break burst.
GRB 090323, GRB 090328, and GRB 090902B do not have clear jet-break characteristics
in their optical band (see Figs. 2, 4, and 6 of Cenko et al. 2011). GRB 090926A
is one of the brightest long bursts detected by the GBM and LAT instruments on Fermi
with high-energy events up to $\sim$ 20 GeV. This burst shows an extra hard component
in its integrated spectrum, whose break energy is around 1.4 GeV. The integrated
spectrum can be well fitted by two Band functions (Ackermann et al. 2011),
so the real $E_{\rm p}$ is confusing.

In summary, we were able to synthesize a
sample of 33 high-quality bursts. All of these data were obtained from previously
published studies. Our complete sample is shown in Table 1, which includes
the following information for each GRB: (1) its name; (2) the redshift; and various
spectral fitting parameters, including (3) the spectral peak energy $E_{\rm p}$
(with corresponding error $\sigma_{E_{\rm p}}$), (4) the low-energy
photon index $\alpha$, (5) the high-energy photon index $\beta$; (6) the
$\gamma$-ray fluence $S_\gamma$ (with error $\sigma_{S_{\gamma}}$);
(7) the observed energy band; and (8) the jet break time $t_{\rm b}$
(with error $\sigma_{t_{\rm b}}$).

With the data listed in Table 1, we calculate the isotropic equivalent
gamma-ray energy ($E_{\gamma,{\rm iso}}$) using
\begin{equation}
E_{\gamma,{\rm iso}}=\frac{4\pi D^{2}_{L}(z)S_\gamma}{(1+z)}K,
\end{equation}
where $S_\gamma$ is the measured gamma-ray fluence,
$D_{L}(z)$ is the luminosity distance at redshift \emph{z}, and $K$
is the $K$-correction factor used to correct the gamma-ray fluence
measured within the observed bandpass (taken to be $1-10^{4}$ keV in this
paper) and shift it into the corresponding bandpass seen in the cosmological
rest frame.

Both $\Lambda$CDM and $R_{\rm h}=ct$ are Friedmann-Robertson-Walker
(FRW) cosmologies, but the former assumes specific constitutents in the density,
written as $\rho=\rho_r+\rho_m+\rho_\Lambda$, where $\rho_r$,
$\rho_m$ and $\rho_\Lambda$ are, respectively, the energy densities
for radiation, matter (both luminous and dark) and the cosmological
constant. These densities are often written in terms of today's critical
density, $\rho_c\equiv 3c^2 H_0^2/8\pi G$, represented as
$\Omega_m\equiv\rho_m/\rho_c$, $\Omega_r\equiv\rho_r/\rho_c$,
and $\Omega_\Lambda\equiv \rho_\Lambda/\rho_c$. In a flat
universe with zero spatial curvature, the total scaled energy
density is $\Omega\equiv\Omega_m+\Omega_r+\Omega_\Lambda=1$.
In $R_{\rm h}=ct$, on the other hand, the only constraint is the
total equation of state $p=w\rho$, where $w=-1/3$. Later in
this paper, we will discuss how these two formulations are related
to each other, particularly how the constraint $w=-1/3$ uniquely forces
$\Omega_m=0.27$ in $\Lambda$CDM when $p_\Lambda=-\rho_\Lambda$
(Melia 2012c).

In $\Lambda$CDM, the luminosity distance is given as
\begin{equation}
D_{L}^{\Lambda {\rm CDM}}(z)={c\over H_{0}}{(1+z)\over\sqrt{\mid\Omega_{k}\mid}}\; sinn\left\{\mid\Omega_{k}\mid^{1/2}
\times\int_{0}^{z}{dz\over\sqrt{(1+z)^{2}(1+\Omega_{m}z)-z(2+z)\Omega_{\Lambda}}}\right\}\;,
\end{equation}
where $c$ is the speed of light, and $H_{0}$ is the Hubble constant at the present time.
In this equation, $\Omega_{k}$ is defined similarly to $\Omega_m$ and represents the
spatial curvature of the Universe---appearing as a term proportional to the spatial
curvature constant $k$ in the Friedmann equation. Also, $sinn$ is $\sinh$ when
$\Omega_{k}>0$ and $\sin$ when $\Omega_{k}<0$. For a flat Universe with
$\Omega_{k}=0$, Equation (2) simplifies to the form $(1+z)c/H_{0}$
times the integral. For the $R_{\rm h}=ct$ Universe, the luminosity distance is
given by the much simpler expression
\begin{equation}
D_{L}^{R_{\rm h}=ct}=\frac{c}{H_{0}}(1+z)\ln(1+z)\;.
\end{equation}
The factor $c/H_0$ is in fact the gravitational horizon $R_{\rm h}(t_0)$ at the
present time, so we may also write the luminosity distance as
\begin{equation}
D_{L}^{R_{\rm h}=ct}=R_{\rm h}(t_0)(1+z)\ln(1+z)\;.
\end{equation}

%\clearpage
\begin{deluxetable}{lllllllll}
%\begin{longtable}{llllllll}
%\tablewidth{400pt}
\tablewidth{500pt}
%\tabletypesize{\tiny}
\tabletypesize{\footnotesize}
\tablecaption{GRB Prompt Emission Parameters and the Jet Break Time}\tablenum{1}
\tablehead{\colhead{GRB}&\colhead{\emph{z}}&\colhead{$E_{\rm p}(\sigma_{E_{\rm p}})$}&\colhead{$\alpha$}&\colhead{$\beta$}
&\colhead{$S_{\gamma}(\sigma_{S_\gamma})$}&\colhead{Band}&\colhead{$t_{\rm b}(\sigma_{t_{\rm b}})$}& \colhead{References}
\\  \colhead{}& \colhead{}&\colhead{(keV)}&\colhead{}&\colhead{}&\colhead{($10^{-6}$ erg cm$^{-2}$)}&\colhead{(keV)}&\colhead{(days)}& \colhead{}
} \startdata
970828	&	0.9578	&	297.7	$\pm$	59.5	&	-0.70	&	-2.07	&	96	$\pm$	9.6	&	20	-	2000	&	2.2	$\pm$	0.4	& 1, 2, 2, 2 \\
980703	&	0.966	&	254	$\pm$	50.8	&	-1.31	&	-2.40	&	22.6	$\pm$	2.3	&	20	-	2000	&	3.4	$\pm$	0.5	& 3, 4, 4, 5 \\
990123	&	1.6	&	780.8	$\pm$	61.9	&	-0.89	&	-2.45	&	300	$\pm$	40	&	40	-	700	&	2.04	$\pm$	0.46	& 6, 7, 7, 6 \\
990510	&	1.62	&	161.5	$\pm$	16.1	&	-1.23	&	-2.70	&	19	$\pm$	2	&	40	-	700	&	1.6	$\pm$	0.2	    & 8, 7, 7, 9 \\
990705	&	0.8424	&	188.8	$\pm$	15.2	&	-1.05	&	-2.20	&	75	$\pm$	8	&	40	-	700	&	1	$\pm$	0.2	    & 2, 2, 2, 2 \\
990712	&	0.43	&	65	$\pm$	11	&	-1.88	&	-2.48	&	6.5	$\pm$	0.3	&	40	-	700	&	1.6	$\pm$	0.2	            & 8, 7, 7, 10 \\
991216	&	1.02	&	317.3	$\pm$	63.4	&	-1.23	&	-2.18	&	194	$\pm$	19	&	20	-	2000	&	1.2	$\pm$	0.4	& 11, 4, 4, 12 \\
000926	&	2.07	&	100	$\pm$	7	&	-1.10	&	-2.43	&	26	$\pm$	4	&	20	-	2000	&	1.74	$\pm$	0.11	& 13, 14, 13, 15 \\
010222	&	1.48	&	291	$\pm$	43	&	-1.05	&	-2.14	&	88.6	$\pm$	1.3	&	40	-	700	&	0.93	$\pm$	0.15	& 16, 17, 17, 16 \\
011211	&	2.14	&	59.2	$\pm$	7.6	&	-0.84	&	-2.30	&	5	$\pm$	0.5	&	40	-	700	&	1.56	$\pm$	0.02	& 18, 19, 18, 20 \\
020124	&	3.2	&	86.9	$\pm$	15	&	-0.79	&	-2.30	&	8.1	$\pm$	0.8	&	2	-	400	&	3	$\pm$	0.4	            & 21, 22, 22, 23 \\
020405	&	0.69	&	192.5	$\pm$	53.8	&	0.00	&	-1.87	&	74	$\pm$	0.7	&	15	-	2000	&	1.67	$\pm$	0.52	& 24, 24, 24, 24 \\
020813	&	1.25	&	142	$\pm$	13	&	-0.94	&	-1.57	&	97.9	$\pm$	10	&	2	-	400	&	0.43	$\pm$	0.06	& 25, 22, 22, 25 \\
021004	&	2.332	&	79.8	$\pm$	30	&	-1.01	&	-2.30	&	2.6	$\pm$	0.6	&	2	-	400	&	4.74	$\pm$	0.14	& 26, 22, 22, 27 \\
021211	&	1.006	&	46.8	$\pm$	5.5	&	-0.86	&	-2.18	&	3.5	$\pm$	0.1	&	2	-	400	&	1.4	$\pm$	0.5	        & 28, 22, 22, 29 \\
030226	&	1.986	&	97	$\pm$	20	&	-0.89	&	-2.30	&	5.61	$\pm$	0.65	&	2	-	400	&	1.04	$\pm$	0.12	& 30, 22, 22, 31 \\
030328	&	1.52	&	126.3	$\pm$	13.5	&	-1.14	&	-2.09	&	37	$\pm$	1.4	&	2	-	400	&	0.8	$\pm$	0.1	    & 32, 22, 22, 33 \\
030329	&	0.1685	&	67.9	$\pm$	2.2	&	-1.26	&	-2.28	&	163	$\pm$	10	&	2	-	400	&	0.5	$\pm$	0.1	        & 34, 22, 22, 35 \\
030429	&	2.6564	&	35	$\pm$	9	&	-1.12	&	-2.30	&	0.85	$\pm$	0.14	&	2	-	400	&	1.77	$\pm$	1	& 36, 22, 22, 37 \\
041006	&	0.716	&	63.4	$\pm$	12.7	&	-1.37	&	-2.30	&	19.9	$\pm$	1.99	&	25	-	100	&	0.16	$\pm$	0.04	& 2, 2, 2, 38\\
050401	&	2.9	&	128	$\pm$	30	&	-1.00	&	-2.45	&	19.3	$\pm$	0.4	&	20	-	2000	&	1.5	$\pm$	0.5	        & 39, 39, 39, 40 \\
050408	&	1.2357	&	19.93	$\pm$	4	&	-1.98	&	-2.30	&	1.9	$\pm$	0.19	&	30	-	400	&	0.28	$\pm$	0.17 & 2, 2, 2, 41 	\\
050416A	&	0.653	&	17	$\pm$	5	&	-1.01	&	-3.40	&	0.35	$\pm$	0.03	&	15	-	150	&	1	$\pm$	0.7	    & 39, 39, 39, 40 \\
050525A	&	0.606	&	79	$\pm$	3.3	&	-0.99	&	-8.84	&	20.1	$\pm$	0.5	&	15	-	350	&	0.28	$\pm$	0.12	& 42, 42, 42, 42 \\
060206	&	4.048	&	75.5	$\pm$	19.4	&	-1.06	&	...	&	0.84	$\pm$	0.04	&	15	-	150	&	2.3	$\pm$	0.11 & 39, 39, 39, 40 \\
060526	&	3.21	&	25	$\pm$	5	&	-1.10	&	-2.20	&	0.49	$\pm$	0.06	&	15	-	150	&	2.77	$\pm$	0.3	& 39, 39, 39, 43 \\
060614	&	0.125	&	49	$\pm$	40	&	-1.00	&	...	&	22	$\pm$	2.2	&	15	-	150	&	1.38	$\pm$	0.04	         & 39, 39, 39, 44 \\
070125	&	1.547	&	367	$\pm$	58	&	-1.10	&	-2.08	&	174	$\pm$	17	&	20	-	10000	&	3.8	$\pm$	0.4	        & 39, 39, 39, 45 \\
071010A	&	0.98	&	16.16	$\pm$	10.6	&	-1.00	&	...	&	0.47	$\pm$	0.11	&	15	-	350	&	0.96	$\pm$	0.09 & 46, 46, 47, 46 \\
071010B	&	0.947	&	52	$\pm$	12	&	-1.25	&	-2.65	&	4.78	$\pm$	2.035	&	20	-	1000	&	3.44	$\pm$	0.39 & 48, 49, 49, 50 \\
090618	&	0.54	&	134	$\pm$	19	&	-1.42	&	...	&	105	$\pm$	1	&	15	-	150	&	0.744	$\pm$	0.07	             & 51, 51, 51, 51 \\
110503A	&	1.61	&	219	$\pm$	20	&	-0.98	&	-2.70	&	26	$\pm$	2	&	20	-	5000	&	2.14	$\pm$	0.21	& 52, 52, 52, 53 \\
110801A	&	1.858	&	140	$\pm$	60	&	-1.70	&	-2.50	&	7.3	$\pm$	1.3	&	15	-	1200	&	1	$\pm$	0.1         & 54, 54, 54, 55 \\
\enddata
\tablenotetext{References:\hskip0.2in}{The references appear in the following order: redshift, spectral parameters, fluence, and break time:
(1) Djorgovski et al. (2001); (2) Wang \& Dai et al. (2006); (3) Djorgovski et al. (1998); (4) Jimenez et al. (2001);
(5) Frail et al. (2003); (6) Kulkarni et al. (1999); (7) Amati et al. (2002); (8) Vreeswijk et al. (2001); (9) Stanek et al. (1999);
(10) Bj$\ddot{o}$rnsson et al. (2001); (11) Djorgovski et al. (1999); (12) Halpern et al. (2000); (13) Amati et al. (2006);
(14) Xiao et al. (2009); (15) Sagar et al. (2001); (16) Galama et al. (2003); (17) Guidorzi et al. (2011);
(18) Holland et al. (2002); (19) Amati (2003); (20) Jakobsson et al. (2003); (21) Hjorth et al. (2003);
(22) Sakamoto et al. (2005); (23) Berger et al. (2002); (24) Price et al. (2003); (25) Barth et al. (2003);
(26) M$\ddot{o}$ller et al. (2002); (27) Holland et al. (2003); (28) Vreeswijk et al. (2003); (29) Holland
et al. (2004); (30) Greiner et al. (2003); (31) Klose et al. (2004); (32) Martini et al. (2003);
(33) Andersen et al. (2003); (34) Bloom et al. (2003b); (35) Berger et al. (2003); (36) Weidinger
et al. (2003); (37) Jakobsson et al. (2004); (38) Stanek et al. (2005); (39) Ghirlanda et al. (2008);
(40) Ghirlanda et al. (2007); (41) Godet et al. (2005); (42) Blustin et al. (2006);  (43) Dai et al. (2007);
(44) Della Valle et al. (2006); (45) Chandra et al. (2008); (46) Covino et al. (2008); (47) Butler et al. (2010);
(48) Amati et al. (2008); (49) Golenetskii et al. (2007); (50) Kann et al. (2007); (51) Schady et al. (2009);
(52) Golenetskii et al. (2011); (53) Kann et al. (2011); (54) Sakamoto et al. (2011); (55) Nicuesa Guelbenzu et al. (2011).
}
\end{deluxetable}

%\clearpage
For each of the GRB sources listed in Table 1, we have derived the equivalent
isotropic energy according to Equation~(1) and listed it in Table 2. Here,
$E_{\gamma,{\rm iso}}^{\Lambda{\rm CDM}}$ and ${E}_{\gamma,{\rm iso}}^{R_{\rm h}
=ct}$ are the isotropic energies in $\Lambda$CDM and $R_{h}=ct$, respectively, remembering
that all of the standard candle features must be re-calibrated for each assumed
expansion scenario. Note, however, that $E'_{\rm p}$ and $t'_{\rm b}$ depend only
on redshift $z$ and are therefore independent of the assumed cosmology.
One notices immediately that, although the numbers differ slightly, they are in
fact remarkably similar, even though $D_L^{\Lambda{\rm CDM}}$ and
$D_L^{R_{\rm h}=ct}$ have quite different formulations. This is another
consequence of the fact that $\Lambda$CDM mimics the $R_{\rm h}=ct$
Universe quite closely, as we will discuss later in this paper (see also
Melia 2012c).

As we alluded to in the introduction, our approach is similar to that presented by
Liang \& Zhang (2005), in which we seek an empirical relationship between
$E_{\gamma,{\rm iso}}$, $E'_{\rm p}$, and $t'_{\rm b}$, known as
the Liang-Zhang relation. Our form of the luminosity correlation
is written as follows:
\begin{equation}
\log E_{\gamma,{\rm iso}}=\kappa_{0}+\kappa_{1}\log E'_{\rm p}+\kappa_{2}\log t'_{\rm b}\;,
\end{equation}
where $E'_{\rm p}=E_{\rm p}(1+z)$ in keV and $t'_{\rm b}=t_{\rm b}/(1+z)$ in days.
To find the best-fit coefficients $\kappa_0$, $\kappa_1$ and $\kappa_2$,
we follow the technique described in D'Agostini (2005). Let us first simplify the
notation by writing $x_{1}=\log E'_{\rm p}$, $x_{2}=\log t'_{\rm b}$, and
$y=\log E_{\gamma,{\rm iso}}$. Then, the joint likelihood function for the
coefficients $\kappa_{0}$, $\kappa_{1}$, $\kappa_{2}$ and the intrinsic
scatter $\sigma_{\rm int}$, is
\begin{equation}
\begin{split}
L(\kappa_{0},\kappa_{1},\kappa_{2},\sigma_{\rm int})\propto\prod_{i}\frac{1}{\sqrt{\sigma^{2}_{\rm int}+\sigma^{2}_{y_{i}}+\kappa^{2}_{1}\sigma^{2}_{x_{1,i}}+\kappa^{2}_{2}\sigma^{2}_{x_{2,i}}}}\times\qquad\qquad\\
\null\qquad\exp\left[-\frac{(y_{i}-\kappa_{0}-\kappa_{1}x_{1,i}-\kappa_{2}x_{2,i})^{2}}{2(\sigma^{2}_{\rm int}+\sigma^{2}_{y_{i}}+\kappa^{2}_{1}\sigma^{2}_{x_{1,i}}+\kappa^{2}_{2}\sigma^{2}_{x_{2,i}})}\right]\;,
\end{split}
\end{equation}
where $i$ is the corresponding serial number of each GRB in our sample.

\clearpage
\begin{deluxetable}{lllll}
\tablewidth{370pt}
%\tabletypesize{\tiny}
\tabletypesize{\footnotesize}
\tablecaption{Derived Rest-frame Burst Properties in $\Lambda$CDM and $R_{\rm h}=ct$}\tablenum{2}
\tablehead{\colhead{GRB}&\colhead{log $E'_{\rm p}(\sigma_{E'_{\rm p}})$}&\colhead{log $t'_{\rm b}(\sigma_{t'_{\rm b}})$}&\colhead{log $E_{\gamma,{\rm iso}}^{\Lambda{\rm CDM}}(\sigma_{E_{\gamma}})$}&\colhead{log ${E}_{\gamma,{\rm iso}}^{R_{\rm h}=ct}(\sigma_{{E}_{\gamma}})$}
\\  \colhead{}&\colhead{(keV)}&\colhead{(days)}& \colhead{(erg)}&\colhead{(erg)}
} \startdata
970828	&	2.77	$\pm$	0.09	&	0.051	$\pm$	0.079	&	53.48	$\pm$	0.04	&	53.39	$\pm$	0.04	\\
980703	&	2.70	$\pm$	0.09	&	0.238	$\pm$	0.064	&	52.85	$\pm$	0.04	&	52.76	$\pm$	0.04	\\
990123	&	3.31	$\pm$	0.03	&	-0.105	$\pm$	0.098	&	54.61	$\pm$	0.06	&	54.51	$\pm$	0.06	\\
990510	&	2.63	$\pm$	0.04	&	-0.214	$\pm$	0.054	&	53.31	$\pm$	0.05	&	53.21	$\pm$	0.05	\\
990705	&	2.54	$\pm$	0.03	&	-0.265	$\pm$	0.087	&	53.41	$\pm$	0.05	&	53.31	$\pm$	0.05	\\
990712	&	1.97	$\pm$	0.07	&	0.049	$\pm$	0.054	&	51.92	$\pm$	0.02	&	51.85	$\pm$	0.02	\\
991216	&	2.81	$\pm$	0.09	&	-0.226	$\pm$	0.145	&	53.84	$\pm$	0.04	&	53.74	$\pm$	0.04	\\
000926	&	2.49	$\pm$	0.03	&	-0.247	$\pm$	0.027	&	53.53	$\pm$	0.07	&	53.44	$\pm$	0.07	\\
010222	&	2.86	$\pm$	0.06	&	-0.426	$\pm$	0.070	&	53.96	$\pm$	0.01	&	53.86	$\pm$	0.01	\\
011211	&	2.27	$\pm$	0.06	&	-0.304	$\pm$	0.006	&	53.01	$\pm$	0.04	&	52.92	$\pm$	0.04	\\
020124	&	2.56	$\pm$	0.07	&	-0.146	$\pm$	0.058	&	53.39	$\pm$	0.04	&	53.32	$\pm$	0.04	\\
020405	&	2.51	$\pm$	0.12	&	-0.005	$\pm$	0.135	&	53.13	$\pm$	0.00	&	53.04	$\pm$	0.00	\\
020813	&	2.50	$\pm$	0.04	&	-0.719	$\pm$	0.061	&	54.14	$\pm$	0.04	&	54.04	$\pm$	0.04	\\
021004	&	2.42	$\pm$	0.16	&	0.153	$\pm$	0.013	&	52.66	$\pm$	0.10	&	52.58	$\pm$	0.10	\\
021211	&	1.97	$\pm$	0.05	&	-0.156	$\pm$	0.155	&	52.14	$\pm$	0.01	&	52.04	$\pm$	0.01	\\
030226	&	2.46	$\pm$	0.09	&	-0.458	$\pm$	0.050	&	52.89	$\pm$	0.05	&	52.80	$\pm$	0.05	\\
030228	&	2.50	$\pm$	0.05	&	-0.498	$\pm$	0.054	&	53.57	$\pm$	0.02	&	53.47	$\pm$	0.02	\\
030329	&	1.90	$\pm$	0.01	&	-0.369	$\pm$	0.087	&	52.20	$\pm$	0.03	&	52.17	$\pm$	0.03	\\
030429	&	2.11	$\pm$	0.11	&	-0.315	$\pm$	0.245	&	52.25	$\pm$	0.07	&	52.17	$\pm$	0.07	\\
041006	&	2.04	$\pm$	0.09	&	-1.030	$\pm$	0.109	&	53.01	$\pm$	0.04	&	52.92	$\pm$	0.04	\\
050401	&	2.70	$\pm$	0.10	&	-0.415	$\pm$	0.145	&	53.63	$\pm$	0.01	&	53.56	$\pm$	0.01	\\
050408	&	1.65	$\pm$	0.09	&	-0.902	$\pm$	0.264	&	52.40	$\pm$	0.04	&	52.30	$\pm$	0.04	\\
050416A	&	1.45	$\pm$	0.13	&	-0.218	$\pm$	0.304	&	50.96	$\pm$	0.04	&	50.87	$\pm$	0.04	\\
050525A	&	2.10	$\pm$	0.02	&	-0.759	$\pm$	0.186	&	52.38	$\pm$	0.01	&	52.29	$\pm$	0.01	\\
060206	&	2.58	$\pm$	0.11	&	-0.341	$\pm$	0.021	&	52.76	$\pm$	0.02	&	52.72	$\pm$	0.02	\\
060526	&	2.02	$\pm$	0.09	&	-0.182	$\pm$	0.047	&	52.42	$\pm$	0.05	&	52.36	$\pm$	0.05	\\
060614	&	1.74	$\pm$	0.35	&	0.089	$\pm$	0.013	&	51.26	$\pm$	0.04	&	51.23	$\pm$	0.04	\\
070125	&	2.97	$\pm$	0.07	&	0.174	$\pm$	0.046	&	53.98	$\pm$	0.04	&	53.88	$\pm$	0.04	\\
071010A	&	1.51	$\pm$	0.28	&	-0.314	$\pm$	0.041	&	51.40	$\pm$	0.10	&	51.30	$\pm$	0.10	\\
071010B	&	2.01	$\pm$	0.10	&	0.247	$\pm$	0.049	&	52.25	$\pm$	0.18	&	52.15	$\pm$	0.18	\\
090618	&	2.31	$\pm$	0.06	&	-0.316	$\pm$	0.041	&	53.34	$\pm$	0.00	&	53.26	$\pm$	0.00	\\
110503A	&	2.76	$\pm$	0.04	&	-0.086	$\pm$	0.043	&	53.27	$\pm$	0.03	&	53.17	$\pm$	0.03	\\
110801A	&	2.60	$\pm$	0.19	&	-0.456	$\pm$	0.043	&	52.99	$\pm$	0.08	&	52.90	$\pm$	0.08	\\
\enddata
%\tablenotetext{a}{In units of $10^{-6}$ erg cm$^{-2}$}
\end{deluxetable}

\clearpage
The best-fit luminosity correlation is shown in Figure~1, together with the
data, for both $\Lambda$CDM (left panel) and $R_{\rm h}=ct$ (right
panel). For this exercise, we assumed a flat $\Lambda$CDM cosmology
with $\Omega_{m}=0.29$ and $H_{0}=69.32$ km s$^{-1}$ Mpc$^{-1}$,
obtained from the 9-yr WMAP data (Bennett et al. 2012). Using the
above optimization method, we find that in $\Lambda$CDM the best-fit
correlation between $E_{\gamma,{\rm iso}}$, and $E'_{\rm p }$ and
$t'_{\rm b}$, is
\begin{equation}
\log E_{\gamma,{\rm iso}}=(48.44\pm0.38)+(1.83\pm0.15)\log E'_{\rm p}
-(0.81\pm0.22)\log t'_{\rm b}\;,
\end{equation}
with an intrinsic scatter $\sigma_{\rm int}=0.25\pm0.06$. (In Table 2,
this energy is labeled $E_{\gamma,{\rm iso}}^{\Lambda{\rm CDM}}$.)
The best-fitting curve is plotted in the left panel of Figure~1.

\vskip0.2in
\begin{figure*}[hp]
\hskip-0.3in
\includegraphics[angle=0,scale=0.65]{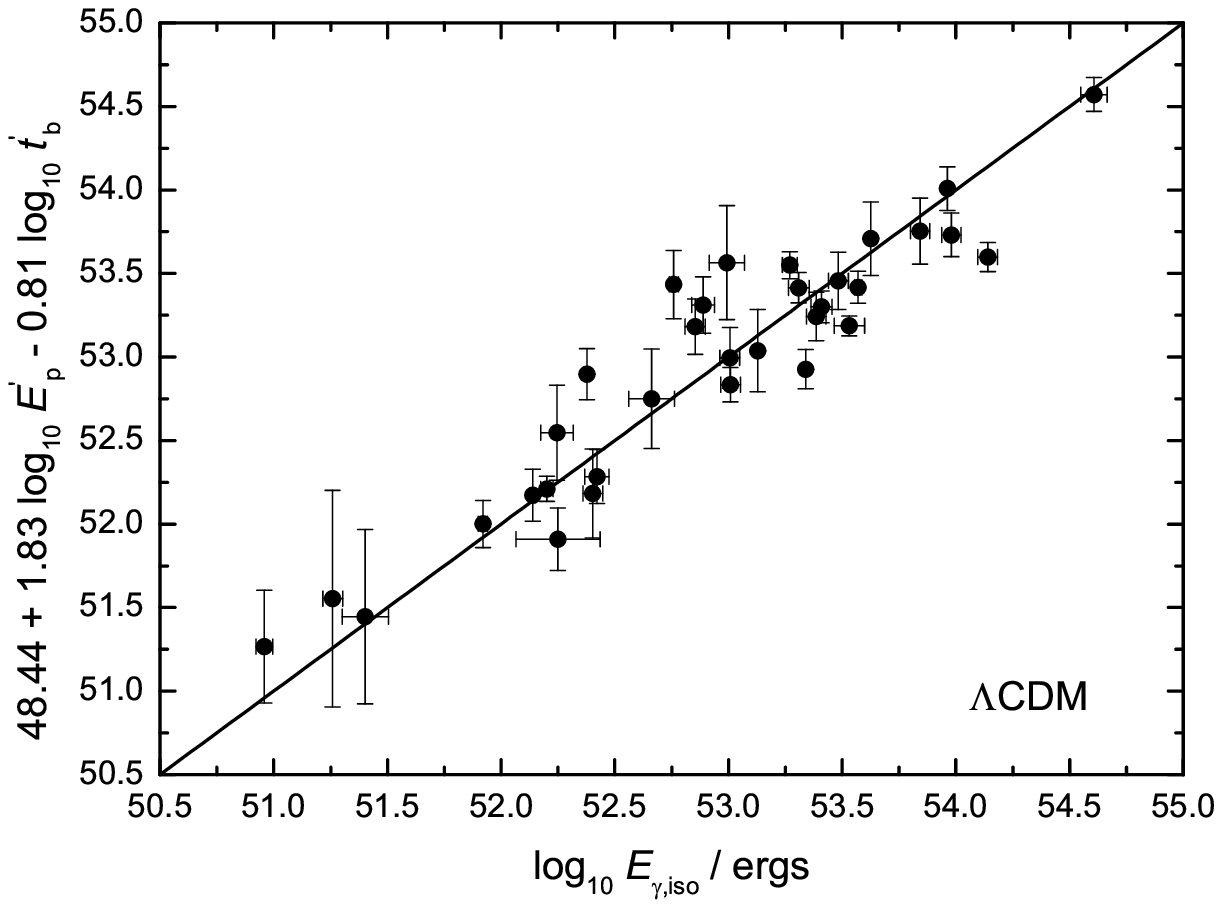}\hskip-0.5in
\includegraphics[angle=0,scale=0.65]{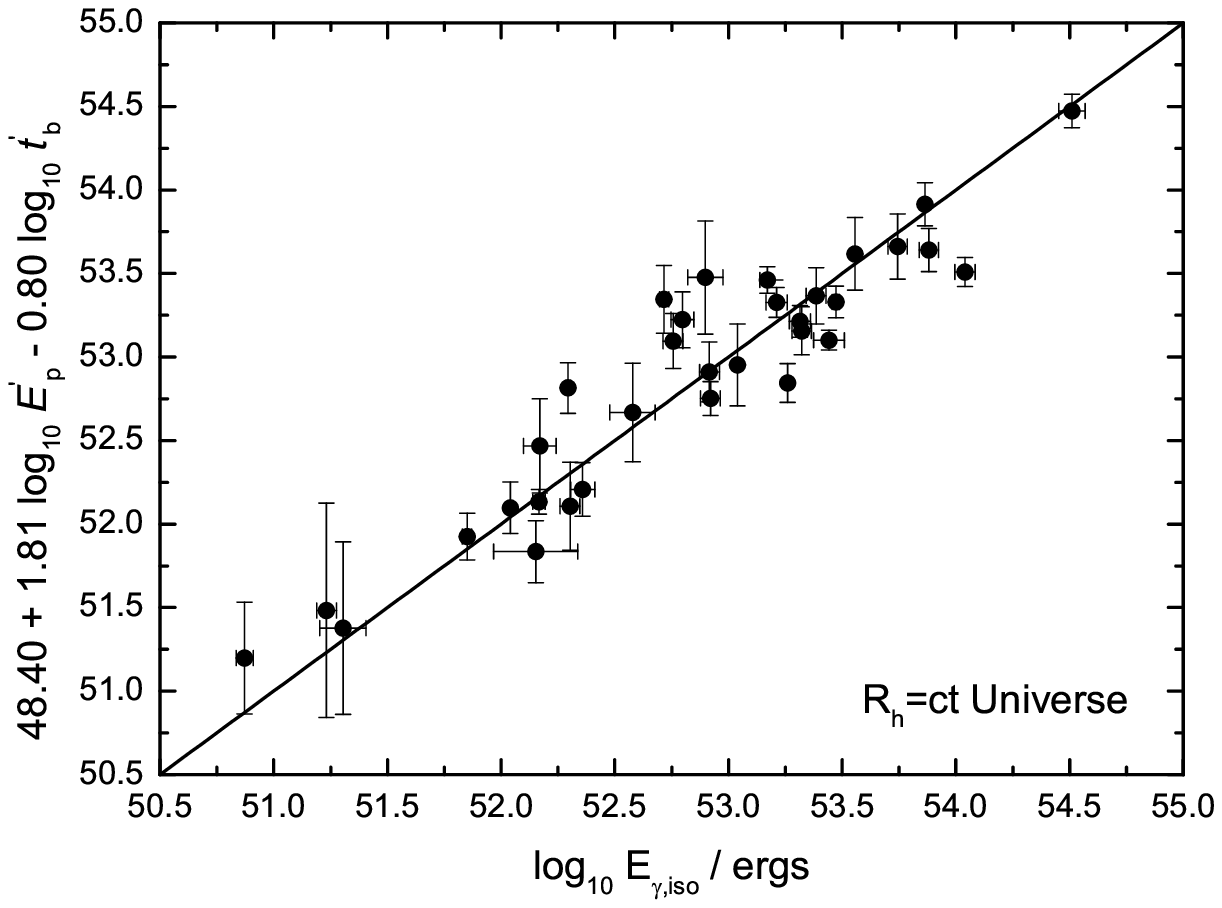}
\caption{The $E_{\gamma,{\rm iso}}$ versus $E'_{\rm p}-t'_{\rm b}$ correlation in
$\Lambda$CDM and $R_{\rm h}=ct$. The solid curves show the best fitting results
from Equation~(5).}\label{correlation}
\end{figure*}

\noindent In the $R_{\rm h}=ct$ Universe, there is only one free parameter---the Hubble
constant $H_{0}$. We note, however, that both the data and the theoretical curves
depend on $1/H_0$, since we do not know the absolute value of the
GRB luminosity. As such, though formally $H_0$ is a free parameter for both
$\Lambda$CDM and the $R_{\rm h}=ct$ Universe, in reality the fits we discuss
in this paper do not depend on its actual value. For the sake of consistency, we
will adopt the standard $H_0=69.32$ km s$^{-1}$ Mpc$^{-1}$ throughout our analysis
and discussion.

Using the above methodology, we find that the best-fitting correlation between
$E_{\gamma,{\rm iso}}$, and $E'_{\rm p}$ and $t'_{\rm b}$, is now
\begin{equation}
\log E_{\gamma,{\rm iso}}=(48.40\pm0.31)+(1.81\pm0.15)\log E'_{\rm p}-(0.80\pm0.22)\log t'_{\rm b}\;,
\end{equation}
with an intrinsic scatter $\sigma_{\rm int}=0.25\pm0.05$. The best-fitting curve is
plotted in the right panel of Figure~1 (and is labeled $E_{\gamma,{\rm iso}}^{R_{\rm h}=ct}$ in Table 2).
These coefficients are quiet similar to those obtained for $\Lambda$CDM.

\section{Optimization of the Model Parameters in $\Lambda$CDM\label{sec:intro}}
The dispersion of the empirical relation for $E_{\gamma,{\rm iso}}$ is so small that it
has served well as a luminosity indicator for cosmology (Liang \& Zhang 2005; Wang
\& Dai 2006). However, since this luminosity indicator is cosmology-dependent,
we cannot use it to constrain the cosmological parameters directly. In order to
avoid circularity issues, we use the following two methods to circumvent
this problem:

\begin{figure*}[hp]
\vskip0.2in
\hskip -0.1in
\includegraphics[angle=0,scale=0.63]{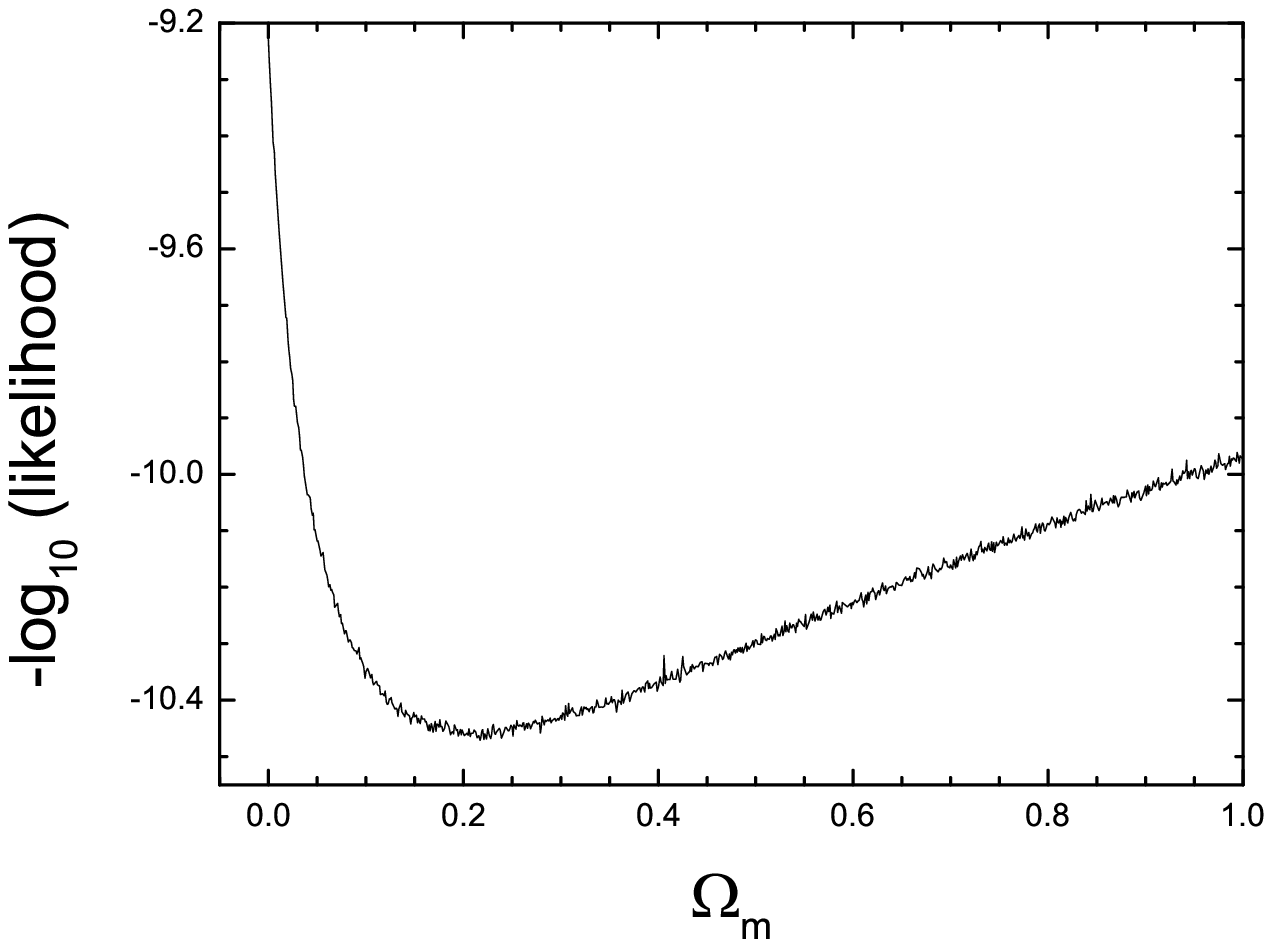}\hskip-0.5in
\includegraphics[angle=0,scale=0.63]{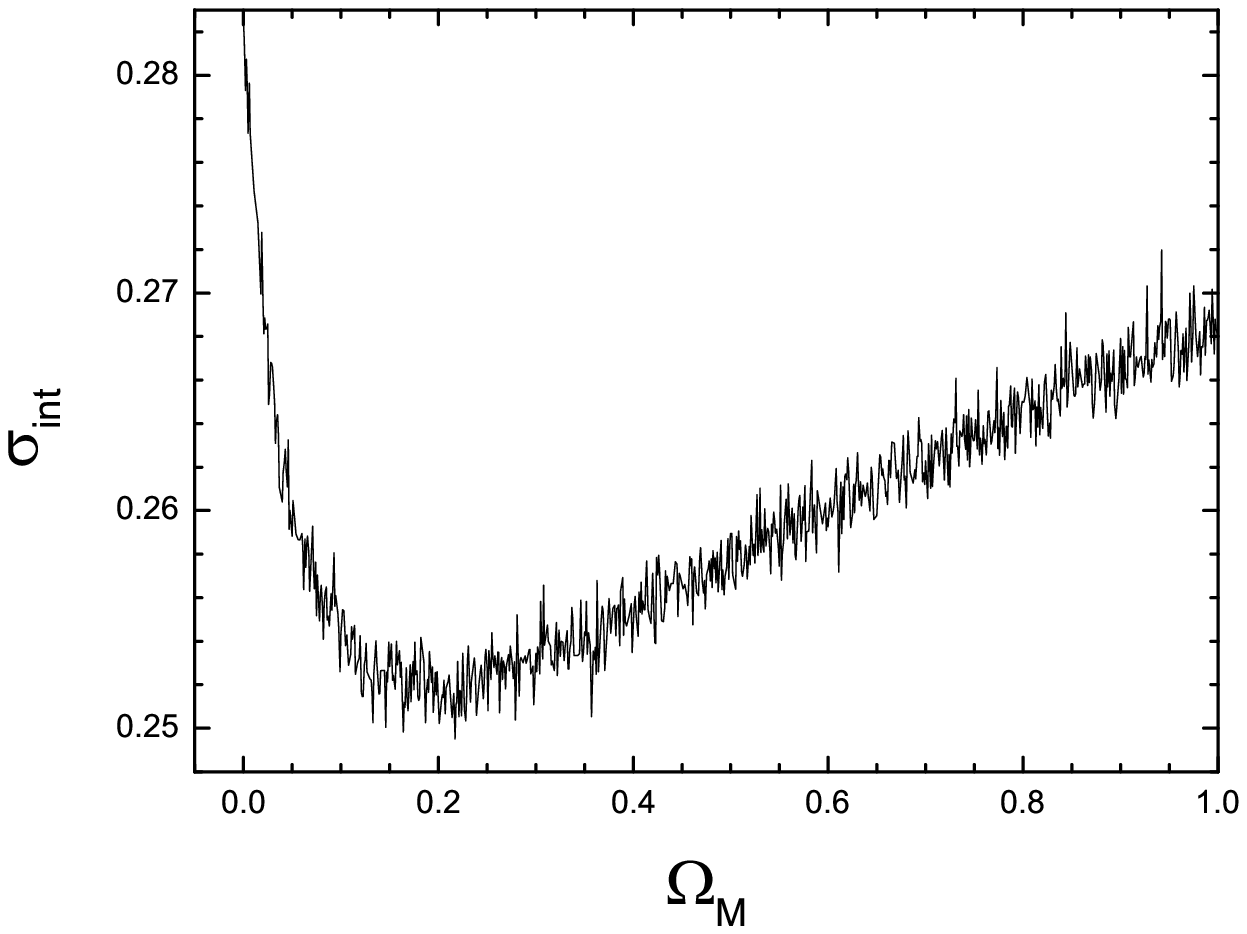}
\caption{Plot of $-\log$(likelihood) (left panel) and the intrinsic scatter
$\sigma_{\rm int}$ (right panel), as functions of $\Omega_{m}$,
obtained by fitting the correlation with the joint likelihood method
in a flat universe.}\label{flat}
\end{figure*}

\emph{Method I}. We repeat the above analysis while varying the cosmological
parameter $\Omega_{m}$, though first under the assumption that the Universe
is flat (Amati et al. 2008; Ghirlanda 2009). The two panels in Figure~2 show
that the values of $-\log$(likelihood) and the intrinsic scatter $\sigma_{\rm int}$
are indeed sensitive to $\Omega_{m}$, showing a clear minimum around
$\Omega_{m}\sim 0.22$. Moreover, the correlation slopes $\kappa_{1}$
and $\kappa_{2}$ are also sensitive to the assumed cosmology, as shown by
the two panels in Figure~3. Using the probability density function,
we can use the joint likelihood method to constrain $\Omega_{m}$ to lie
within the range $0.17-0.75$ at the $1\sigma$ confidence level.

\begin{figure*}[hp]
\vskip0.1in
\hskip-0.1in
\includegraphics[angle=0,scale=0.63]{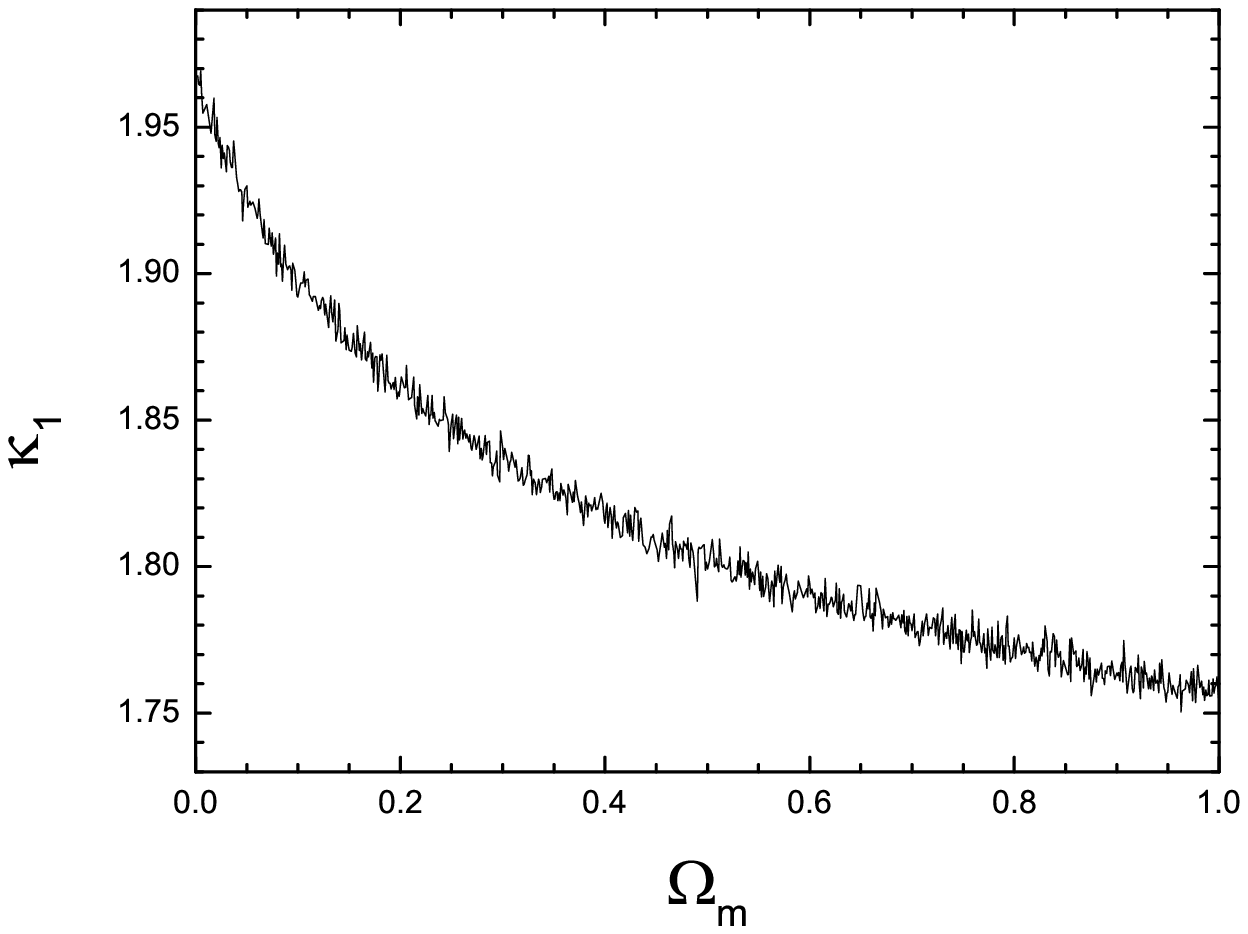}\hskip-0.5in
\includegraphics[angle=0,scale=0.63]{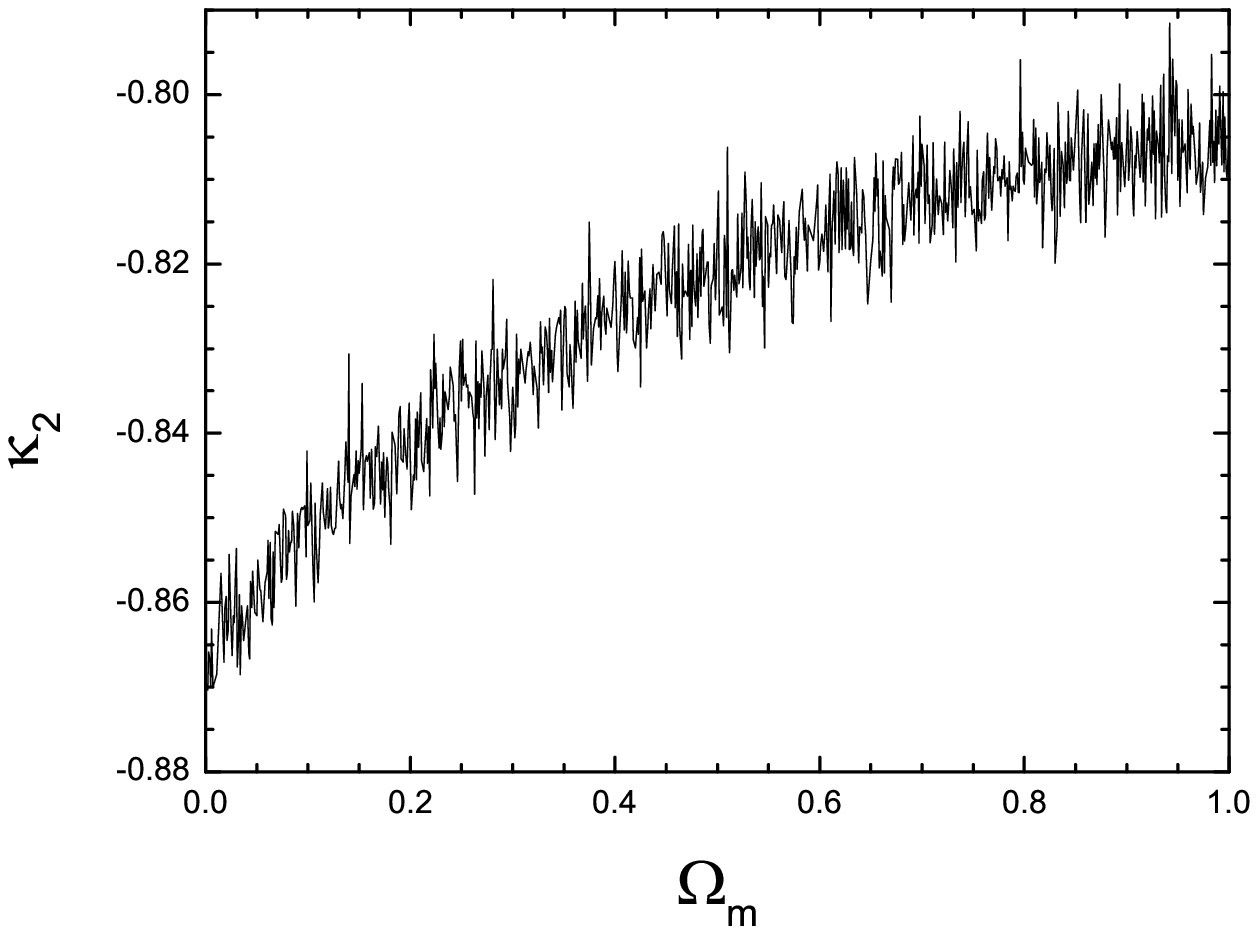}
\caption{The slopes $\kappa_{1}$ (left panel) and $\kappa_{2}$ (right panel)
as functions of $\Omega_{m}$, obtained by fitting the correlation with the
joint likelihood method in a flat universe.}\label{k}
\end{figure*}

If we release the flat universe constraint and allow $\Omega_{m}$ and $\Omega_{\Lambda}$
to vary independently (see Figure~4), the contours show that $\Omega_{m}$ and
$\Omega_{\Lambda}$ are poorly constrained; only an upper limit of $\sim0.68$ and $\sim0.95$
can be set at $1\sigma$ for $\Omega_{m}$ and $\Omega_{\Lambda}$. However,
if we consider only a flat universe, the allowed region at the $1\sigma$
level is restricted by the flat Universe ({\it dashed}) line and the $1\sigma$ contour,
for which $0.10<\Omega_{m}<0.45$ and $0.55<\Omega_{\Lambda}<0.90$. The most probable
values of $\Omega_{m}$ and $\Omega_{\Lambda}$ are $(0.22_{-0.12}^{+0.23}, 0.78_{-0.23}^{+0.12})$.

\begin{figure}[hp]
\centerline{\includegraphics[angle=0,scale=0.7]{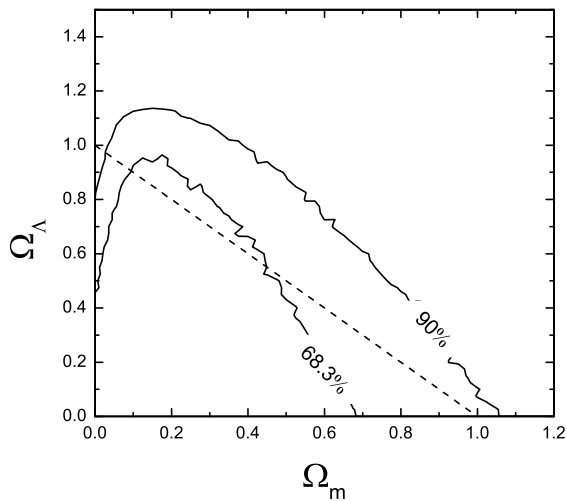}}
\vskip-0.2in
\caption{Contour confidence levels of $\Omega_{m}$ and $\Omega_{\Lambda}$, obtained by fitting
the correlation with Method I.}\label{cont1}
\end{figure}

\emph{Method II}. The distance modulus of a GRB is defined as
\begin{equation}
\mu\equiv5 \log(D_{L}/10 pc)\;,
\end{equation}
in terms of the luminosity distance $D_L$. Using the Liang-Zhang relation, we can recast this
in the form
\begin{equation}
\hat{\mu}=2.5[\kappa_{0}+\kappa_{1} \log E'_{\rm p}+\kappa_{2} \log t'_{\rm b}-\log(4\pi S_{\gamma}K)
+\log(1+z)]-97.45\;.
\end{equation}
However, since the luminosity correlation is cosmology-dependent, $\hat{\mu}$ also depends
on the adopted expansion scenario. We use the following approach to circumvent this
difficulty (see also Liang \& Zhang 2005). This procedure is based on the calculation of the
probability function for a given set of cosmological parameters (denoted by $\bar{\Omega}$, which
includes both $\Omega_{m}$ and $\Omega_{\Lambda}$):

{\it Step 1}. For a given cosmological model, we calibrate and weight the luminosity indicator
corresponding to each choice of parameters $\bar{\Omega}$. In each case, we calculate
the correlation $\hat{E}_{\gamma,{\rm iso}}(\bar{\Omega}; E'_{\rm p}, t'_{\rm b})$, and
evaluate the probability [$w(\bar{\Omega})$] of this relation being the optimal
cosmology-independent luminosity indicator via $\chi^{2}$ statistics, i.e.,
\begin{equation}
\chi_{w}^{2}(\bar{\Omega})=\sum_{i}^{N}\frac{[\log \hat{E}_{\gamma,{\rm iso}}^{i}(\bar{\Omega})
-\log E_{\gamma,{\rm iso}}^{i}(\bar{\Omega})]^{2}}{\sigma_{\log \hat{E}_{\gamma,{\rm iso}}^{i}}^{2}(\bar{\Omega})}\;.
\end{equation}
The probability is then
\begin{equation}
w(\bar{\Omega})\propto e^{-\chi_{w}^{2}(\bar{\Omega})/2}\;.
\end{equation}

{\it Step 2}. We regard the correlation derived for each set of parameters as a cosmology-independent
luminosity indicator without considering its systematic error, and calculate the distance modulus
$\hat{\mu}(\bar{\Omega})$ and its error $\sigma_{\hat{\mu}}$, given by
\begin{equation}
\sigma_{\hat{\mu}_{i}}=\frac{2.5}{\ln 10}\left[\left(\kappa_{1}\frac{\sigma_{E'_{{\rm p},i}}}{E'_{{\rm p},i}}\right)^{2}+
\left(\kappa_{2}\frac{\sigma_{t'_{{\rm b},i}}}{t'_{{\rm b},i}}\right)^{2}+\left(\frac{\sigma_{S_{\gamma,i}}}{S_{\gamma,i}}\right)^{2}
+\left(\frac{\sigma_{K_{i}}}{K_{i}}\right)^{2}+\left(\frac{\sigma_{z_{i}}}{1+z_{i}}\right)^{2}\right]^{1/2}.
\end{equation}
Since both $(\sigma_{K_{i}}/K_{i})^{2}$ and $[\sigma_{z_{i}}/(1+z_{i})]^{2}$ are significantly
smaller than the other terms in Equation~(13), we ignore them in our calculations.

{\it Step 3}. We calculate the theoretical distance modulus $\mu(\Omega)$ for a set of
cosmological parameters (denoted by $\Omega$), and then obtain $\chi^{2}$ from a
comparison of $\mu(\Omega)$ with $\hat{\mu}(\Omega)$, i.e.,
\begin{equation}
\chi^{2}(\bar{\Omega}\mid\Omega)=\sum_{i}^{N}\frac{[\hat{\mu}_{i}(\bar{\Omega})-
\mu_{i}(\Omega)]^{2}}{\sigma_{\hat{\mu}_{i}}^{2}(\bar{\Omega})}\;.
\end{equation}

{\it Step 4}. We then calculate the probability that the cosmological parameter set $\Omega$ is
the correct one according to the luminosity indicator derived from the cosmological
parameter set $\bar{\Omega}$, i.e., we calculate
\begin{equation}
p(\bar{\Omega}\mid\Omega)\propto e^{-\chi^{2}(\bar{\Omega}\mid\Omega)/2}\;.
\end{equation}

\begin{figure}[h]
\vskip 0.2in
\centerline{\includegraphics[angle=0,scale=0.75]{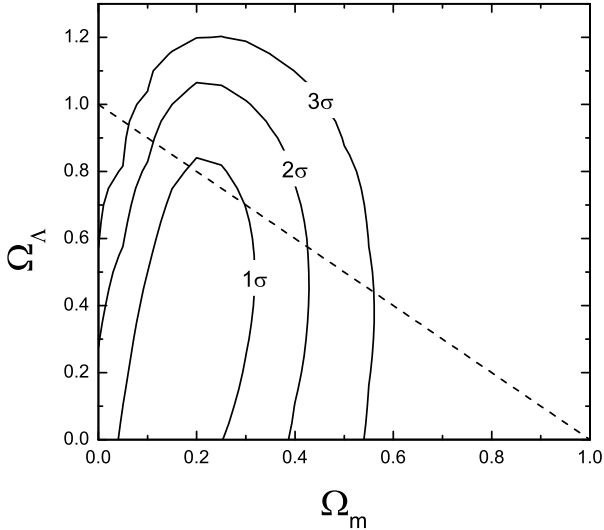}}
\caption{Contour confidence levels of $\Omega_{m}$ and $\Omega_{\Lambda}$, inferred from the current
GRB sample, using method II.}\label{cont2}
\end{figure}

{\it Step 5}. Finally, we integrate $\bar{\Omega}$ over the full cosmological parameter space to
get the final normalized probability that the cosmological parameter set $\Omega$ is the
correct one, i.e.,
\begin{equation}
p(\Omega)=\frac{\int_{\bar{\Omega}}w(\bar{\Omega})p(\bar{\Omega}\mid\Omega)\;d\bar{\Omega}}
{\int_{\bar{\Omega}}w(\bar{\Omega})\;d\bar{\Omega}}\;.
\end{equation}

Figure~5 shows the $1\sigma$ to $3\sigma$ contours of the probability in the
($\Omega_{m}$, $\Omega_{\Lambda}$) plane. The contours show that at the $1\sigma$ level,
$0.04<\Omega_{m}<0.32$, but $\Omega_{\Lambda}$ is poorly constrained; only an upper limit
of $\sim0.84$ can be set at this confidence level. However, if we consider only a flat Universe,
then the allowed range of parameter space is limited by the flat Universe {\it dashed} line and
the $1\sigma$ contour, for which $0.19<\Omega_{m}<0.30$ and $0.7<\Omega_{\Lambda}
<0.81$. The best fit values are $(\Omega_{m},\Omega_{\Lambda})=(0.25_{-0.06}^{+0.05}, 0.75_{-0.05}^{+0.06})$.

\section{The $R_{\rm h}=ct$ Universe}
In the previous section, we considered how the currently available sample of GRB events
with spectral and lightcurve characteristics appropriate for cosmological work may be used
to constrain the principal parameters of the standard model. The
$R_{\rm h}=ct$ Universe, on the other hand, has only one free parameter---the Hubble
constant $H_0$. However, as we have already noted, none of the results presented in this
paper depend on this constant, since $1/H_0$ enters into the determination of both the data
and the theoretical curves. There is therefore no need to reproduce the kind of parameter
optimization for $R_{\rm h}=ct$ that was carried out for $\Lambda$CDM in \S~3. But before
we proceed to compare the Hubble diagrams for the $R_{\rm h}=ct$ Universe and the
optimized $\Lambda$CDM model, we will first briefly summarize the $R_{\rm h}=ct$
cosmology, which is not yet as well known as $\Lambda$CDM.

One may look at the expansion of the Universe in several ways. From the perspective
of the standard model, one guesses the constituents and their equation of state and
then solves the dynamical equations to determine the expansion rate as a function of
time. The second is to use symmetry arguments and our knowledge of the properties
of a gravitational horizon in general relativity (GR) to determine the spacetime curvature,
and thereby the expansion rate, strictly from just the value of the total energy density
$\rho$ and the implied geometry, without necessarily having to worry about the specifics
of the constituents that make up the density itself. This is the approach adopted by
$R_{\rm h}=ct$. In other words, what matters is $\rho$ and the overall equation of
state $p=w\rho$, in terms of the total pressure $p$ and total energy density $\rho$.
In $\Lambda$CDM, one assumes $\rho=\rho_m+\rho_r+\rho_{de}$, i.e., that the
principal constituents are matter, radiation, and an unknown dark energy, and then
infers $w$ from the equations of state assigned to each of these constituents. In
$R_{\rm h}=ct$, it is the aforementioned symmetries and other constraints from
GR that uniquely fix $w$.

Both $\Lambda$CDM and $R_{\rm h}=ct$ are Friedmann-Robertson-Walker (FRW)
cosmologies, but in the latter, Weyl's postulate takes on a more important role
than has been considered before (Melia \& Shevchuk 2012). There is no modification
to GR, and the Cosmological principle is adopted from the start, just like any other
FRW cosmology. However, Weyl's postulate adds a very important ingredient. Most
workers assume that Weyl's postulate is already incorporated into all FRW metrics,
but actually it is only partially incorporated. Simply stated, Weyl's postulate says
that any proper distance $R(t)$ must be the product of a universal expansion
factor $a(t)$ and an unchanging co-moving radius $r$, such that
$R(t)=a(t)r$. The conventional way of writing an FRW metric adopts this
coordinate definition, along with the cosmic time $t$. But what is often
overlooked is the fact that the gravitational radius, $R_{\rm h}$ (see
Equation~4), which has the same definition as the Schwarzschild radius, and
actually coincides with the better known Hubble radius, is in fact itself a
proper distance too (see also Melia \& Abdelqader 2009).  And when
one forces this radius to comply with Weyl's postulate, there is only one
possible choice for $a(t)$, i.e., $a(t)=(t/t_0)$, where $t_0$ is the current
age of the Universe. This also leads to the result that the gravitational radius
must be receding from us at speed $c$, which is in fact how the Hubble radius
was defined in the first place, even before it was recognized as another
manifestation of the gravitational horizon.

The principal difference between $\Lambda$CDM and $R_{\rm h}=ct$ is
how they handle $\rho$ and $p$. In the $R_{\rm h}=ct$ cosmology, the
fact that $a(t)\propto t$ requires that the total pressure $p$ be given as
$p=-\rho/3$. The consequence of this is that quantities such as the
luminosity distance and the redshift dependence of the Hubble constant
$H$, take on very simple, analytical forms (as we have already seen in
Equation~4). Though we won't need it here, we also mention that
the evolution of $H(z)$ in the $R_{\rm h}=ct$ Universe goes as
\begin{equation}
H(z)=H_0(1+z)\;,
\end{equation}
another very simple and elegant expression that is not available in
$\Lambda$CDM. Here, $z$ is the redshift, $R_h=c/H$, and $H_0$ is
the value of the Hubble constant today. These relations are clearly very
relevant to a proper examination of other cosmological observations, and
we are in the process of applying them accordingly. For example, we have
recently demonstrated that the model-independent cosmic chronometer
data (see, e.g., Moresco et al. 2012) are a better match to
$R_{\rm h}=ct$ (using Eq. 17), than the concordance, best-fit
$\Lambda$CDM model (Melia \& Maier 2013).

In the end, regardless of how $\Lambda$CDM or $R_{\rm h}=ct$ handle
$\rho$ and $p$, they must both account for the same cosmological
data. There is growing evidence that, with its empirical approach,
$\Lambda$CDM can function as a reasonable approximation to
$R_{\rm h}=ct$ in some restricted redshift ranges, but apparently
does poorly in others. For example, in using the ansatz
$\rho=\rho_m+\rho_r+\rho_{de}$ to fit the data, one finds that the
$\Lambda$CDM parameters must have
quite specific values, such as $\Omega_m\equiv \rho_m/\rho_c=0.27$
and $w_{de}=-1$, where $\rho_c$ is the critical density and $w_{de}$
is the equation-of-state parameter for dark energy. This is quite
telling because with these parameters, $\Lambda$CDM then requires
$R_{\rm h}(t_0)=ct_0$ today. That is, the best-fit $\Lambda$CDM
parameters describe a universal expansion equal to what it would
have been with $R_{\rm h}=ct$ all along. Other indicators support
the view that using $\Lambda$CDM to fit the data therefore
produces a cosmology almost (but not entirely) identical to
$R_{\rm h}=ct$ (see Melia 2012c).

As we shall see below, the results of our analysis of the GRB HD
produce very similar conclusions to these, i.e., that even though
the internal structure of $\Lambda$CDM would appear to be quite
different from that in $R_{\rm h}=ct$ (compare Equations~2 and
4), in the end, the best fit $\Lambda$CDM model essentially
mimics the universal expansion implied by $R_{\rm h}=ct$.

\section{The GRB Hubble Diagram\label{sec:intro}}
In $\Lambda$CDM, the luminosity indicator, and therefore also the distance modulus
$\mu$, depends on the specific choice of parameter values (for $\Omega_m$ and
$\Omega_\Lambda$). To directly
compare the HD for $\Lambda$CDM with that for $R_{\rm h}=ct$,
we will calculate $\mu$ and $\sigma_{\mu}$ using the best-fit model,
for which $\Omega_{m}=0.25$ and $\Omega_{\Lambda}=0.75$.
The data and best-fit curve are shown together in the left-hand panel
of Figure~6. The $\chi^{2}$ for this fit is calculated according to
\begin{equation}
\chi^{2}=\sum_{i}^{N}\frac{(\mu_{i}^{\rm obs}-\mu_{i}^{\rm th})^{2}}{\sigma_{\rm int}^{2}+\sigma_{\mu_{i}}^{2}}\;,
\end{equation}
where $\mu^{\rm th}$ is theoretical value of the distance modulus, and $\mu^{\rm obs}$ is measured
using the Liang-Zhang relation. Also, $\sigma_{\rm int}$ is the intrinsic scatter obtained from the joint likelihood
analysis, and $\sigma_{\mu}$ is the error for each realization $\mu^{\rm obs}$ of the \emph{N} data points.
Ignoring $H_0$, which does not affect any of these fits, the optimized $\Lambda$CDM model has
two remaining (principal) parameters, $\Omega_m$ and $\Omega_\Lambda$, so with
33 data points, the reduced $\chi^2$ per degree of freedom is \textbf{$\chi^{2}_{\rm dof}=70.07/31=2.26$}.

\begin{figure*}[h]
\hskip-0.3in
\includegraphics[angle=0,scale=0.65]{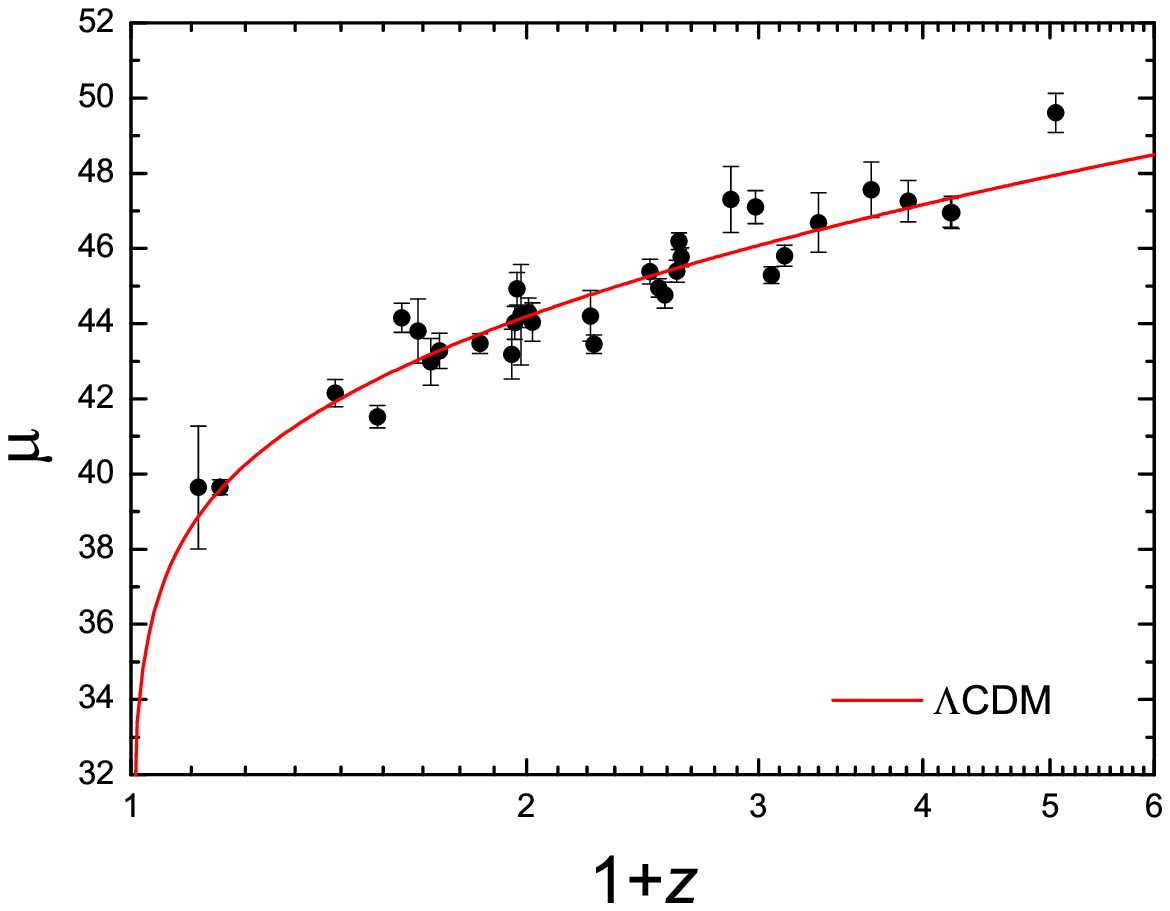}\hskip-0.5in
\includegraphics[angle=0,scale=0.65]{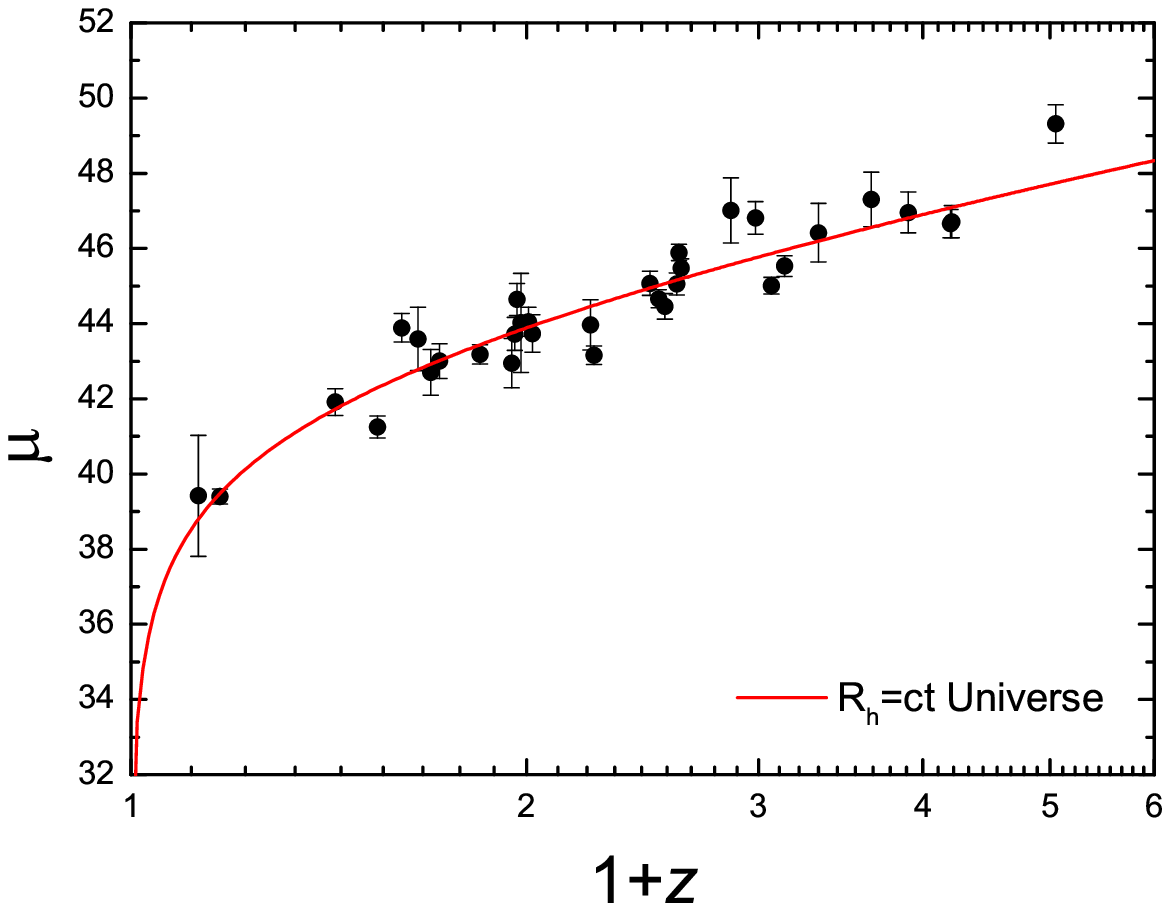}
\caption{Left: Hubble diagram for the GRB sample. The solid curve represents the theoretical $\mu$
in the $\Lambda$CDM model. Right: Same as the left panel, except now the solid curve represents the $R_{h}=ct$ Universe.}\label{HB}
\end{figure*}

The plot actually gives the impression that the fit is better than this $\chi^2_{\rm dof}$ would
suggest. A closer inspection reveals that 5 data points lie more than $\sim 2\sigma$ away from
the best-fit curve. Removal of these data reduces the $\chi^2$ considerably, and may be an
indication that either they are true outliers, or that the errors and intrinsic scatter are greately
underestimated.

The Hubble Diagram for the $R_{\rm h}=ct$ Universe is shown in the right-hand panel of
Figure~6. Both the data and the best-fit curve were calibrated using the expansion implied
by this cosmology (see column 5 in Table 2, and Equation~8). A Hubble constant $H_{0}=69.32$
km s$^{-1}$ Mpc$^{-1}$ was selected to construct the plot, though it has no bearing on
the quality of the fit itself. In this case, since we are ignoring $H_0$ in producing the fit,
there are no remaining free parameters, and the reduced $\chi^2$ per degree of
freedom in $R_{\rm h}=ct$ is $\chi^{2}_{\rm dof}=70.53/33=2.14$. Strictly
based on their $\chi^2_{\rm dof}$'s, the two fits are comparable, though some
concern ought to be expressed about the possible contamination of the GRB sample
by outliers and/or the underestimation of errors and intrinsic scatter. To facilitate a
direct comparison, these two Hubble Diagrams are also shown side by side in Figure~7.

\begin{figure}[h]
\vskip 0.2in
\centerline{\includegraphics[angle=0,scale=1.0]{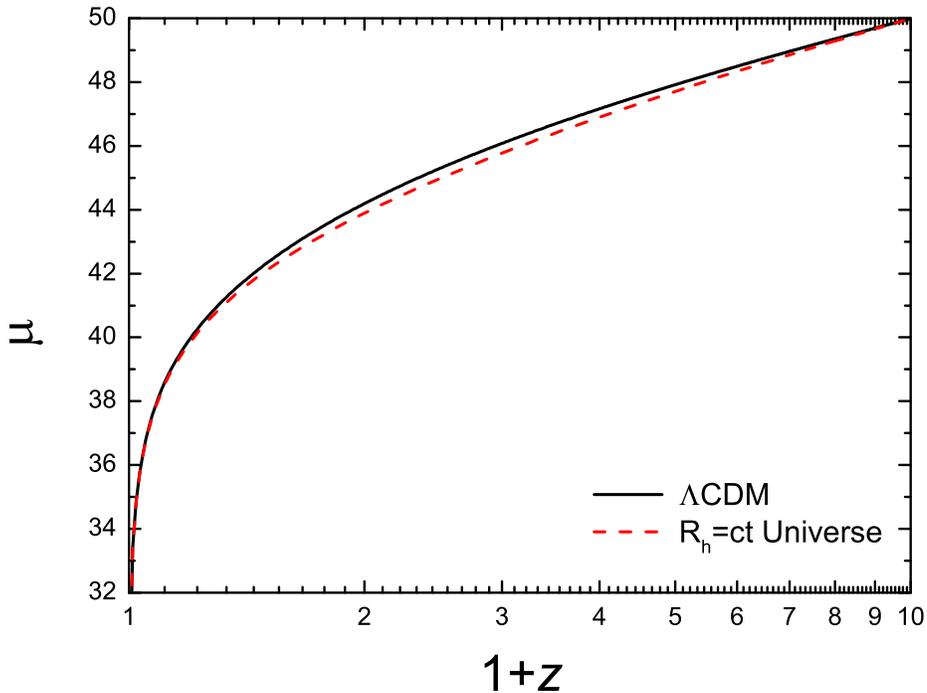}}
\caption{A side-by-side comparison of the theoretical curves in
$\Lambda$CDM and the $R_{h}=ct$ Universe.}\label{comparison}
\end{figure}

To determine the likelikhood of either $R_{\rm h}=ct$ or $\Lambda$CDM
being closer to the ``correct" model, we use the model selection criteria
discussed extensively in Melia \& Maier (2013). For such purposes, the Akaike
Information Criterion (AIC) has become quite common in cosmology (see, e.g.,
Liddle 2004, 2007; Tan \& Biswas 2012). The AIC prefers models with
few parameters to those with many, unless the latter provide a
substantially better fit to the data. This avoids the possibility that by
using a greater number of parameters, one may simply be fitting the noise.

For each fitted model, the AIC is given by
\begin{equation}
\label{eq:AIC}
  {\rm AIC} = \chi^2 + 2\,k\;,
\end{equation}
where $k$ is the number of free parameters. If there are two models
for the data, $\mathcal{M}_1$ and $\mathcal{M}_2$, and they have been
separately fitted, the one with the least resulting AIC is assessed as the
one more likely to be ``true.''  A more quantitative ranking of models can be computed as
follows.  If ${\rm AIC}_\alpha$ comes from model~$\mathcal{M}_\alpha$, the
unnormalized confidence that $\mathcal{M}_\alpha$~is true is the ``Akaike
weight'' $\exp(-{\rm AIC}_\alpha/2)$. Informally, $\mathcal{M}_\alpha$~has
likelihood
\begin{equation}
\label{eq:lastAIC}
{\cal L}(\mathcal{M}_\alpha)=
\frac{\exp(-{\rm AIC}_\alpha/2)}
{\exp(-{\rm AIC}_1/2)+\exp(-{\rm AIC}_2/2)}
\end{equation}
of being closer to the correct model. Thus, the difference ${\rm AIC}_2\nobreak-{\rm
  AIC}_1$ determines the extent to which $\mathcal{M}_1$ is favored
over~$\mathcal{M}_2$.

The choice of proportionality constant (i.e.,~$2$) for $k$ is not entirely arbitrary,
being based on an argument from information theory that has close ties to statistical
mechanics. (More details may be found in Melia \& Maier 2013.)  It is known that the
AIC is increasingly accurate when the number of data points $N$~is large. However,
in all cases, the magnitude of the difference $\Delta=\allowbreak {\rm AIC}_2 -\nobreak
{\rm AIC}_1$ provides a numerical assessment of the evidence that
model~1 is to be preferred over model~2.  A~rule of thumb used in the
literature is that if $\Delta\la2$, the evidence is weak; if $\Delta\approx3$ or~$4$,
it is mildly strong; and if $\Delta\ga5$, it is quite strong.

Several alternatives to the AIC have been considered in the literature, but all
are based on similar arguments. A lesser known one, called the Kullback
Information Criterion (KIC), takes into account the fact that the PDF's of the
various competing models may not be symmetric. The unbiased estimator
for the symmetrized version (Cavanaugh 1999) is given by
\begin{equation}
\label{eq:KIC}
  {\rm KIC} = \chi^2 + 3\,k\;,
\end{equation}
very similar to the AIC, but clearly strengthening the dependence on the
number of free parameters (from $2k$ to $3k$). The rule of thumb
concerning the strength of the evidence in KIC favoring one model over
another is similar to that for AIC, and the likelihood is calculated using
the same Equation~(20), though with AIC$_\alpha$ replaced with
KIC$_\alpha$.

A better known alternative to the AIC is the Bayes Information Criterion
(BIC), an asymptotic ($N\to\infty$) approximation to the outcome of a
conventional Bayesian inference procedure for deciding between models
(Schwarz 1978). This criterion is defined by
\begin{equation}
\label{eq:BIC}
  {\rm BIC} = \chi^2 + (\ln N)\,k,
\end{equation}
and suppresses overfitting very strongly if $N$~is large. This criterion
has already been used by Shi et al. (2012) to compare cosmological models.
In this case, the evidence favoring one model over
another is judged to be positive for a range of $\Delta\equiv {\rm BIC}_1-
{\rm BIC}_2$ between 2 and 6, and is ``strong" for values greater than this.

With the optimized fits we have obtained above, these three model selection
criteria have the following values:
for $R_{\rm h}=ct$, ${\rm AIC}_1=70.53$, ${\rm KIC}_1=70.53$, and
${\rm BIC}_1=70.53$. Whereas, for $\Lambda$CDM, we get
${\rm AIC}_2=74.07$, ${\rm KIC}_2=76.07$, and ${\rm BIC}_2=77.06$.
Therefore, using the AIC, one finds that the likelihood  of $R_{\rm h}=ct$
being closer to the correct cosmology is $85.4\%$,
compared to only $14.6\%$ for
$\Lambda$CDM. The difference is larger using the other
criteria, which show that the $R_{\rm h}=ct$ Universe is favored
over $\Lambda$CDM with a likelihood of $94.1\%$ versus
$5.9\%$ using KIC, and $96.3\%$ versus $3.7\%$ using BIC.
In showing the results of all three criteria, our principal goal
is not so much to dwell on which of these may or may not reflect the
importance of free parameters but, rather, to demonstrate a
universally consistent outcome among the most commonly used
model-selection tools in the literature.
Clearly, the GRB Hubble Diagram favors $R_{\rm h}=ct$
over $\Lambda$CDM. Interestingly, these likelihoods are
very similar to those inferred from our analysis of the cosmic
chronometer data (Melia \& Maier 2013), which showed that
on the basis of those data, the $R_{\rm h}=ct$ Universe
is favored over $\Lambda$CDM with a likelihood of
$82-91\%$ versus $9-18\%$, for these three
model selection criteria.

\section{Discussion and Conclusions\label{sec:intro}}
In this paper, we have added some support to the argument that GRBs
may eventually be used to carry out stringent tests on various cosmological
models. Earlier work on this proposal had indicated that the spectral and
lightcurve features most likely to provide a reliable luminosity indicator
are the peak energy and a proxy for the jet opening angle, which we
have taken to be the time at which a break in the light curve is observed.
In this paper, we have confirmed the notion advanced previously that
examining correlations among these data can indeed produce a luminosity
indicator with sufficient reliability to study the expansion of the Universe.

A notable result of our work, based on the most up-to-date GRB data,
is that a careful statistical analysis of these correlations and their
optimization points to best-fit parameter values in $\Lambda$CDM
remarkably close to those associated with the concordance model.
We have found that the $\Lambda$CDM model most consistent with
the GRB Hubble Diagram has $\Omega_m\approx 0.25$ and
$\Omega_\Lambda\approx 0.75$. In the concordance model,
these values are, respectively, $\approx 0.29$ and $\approx 0.71$
(Hinshaw et al. 2012).

However, for $\Lambda$CDM the reduced $\chi^2_{\rm dof}$ is at best approximately 2.26.
A close inspection of the GRB HD for this model reveals that about $20\%$ of the
data points lie at least $2\sigma$ away from the best-fit curve. This may be
an indication that some contamination of the GRB sample is unavoidable, and
that pure luminosity indicators may never be found for these sources. Of course,
it could also mean that we simply have not yet found the ideal correlation
function, and/or have not yet identified the correct spectral and lightcurve
features to use for this purpose. On the other hand, it could also mean that
we are understimating the errors and intrinsic scatter associated with the
data. Additional work is required in order to better identify the likely resolution
to this problem.

A second principal result of our analysis is that, based on fits to the
GRB HD, the $R_{\rm h}=ct$ Universe is more likely to be closer to the
``correct" model than the optimized $\Lambda$CDM.
One of our goals with this work was to demonstrate the dependence of the
data acquisition on the pre-assumed cosmological model. This appears to be
an unavoidable problem with all cosmological data, except perhaps for the
cosmic chronometer measurements which are obtained independently of
any integrated quantity (such as the luminosity distance) that requires
pre-knowledge of the Universe's expansion history. By calibrating the
GRB data separately for $\Lambda$CDM and $R_{\rm h}=ct$, we have
produced a meaningful side-by-side comparison between these two
cosmologies, showing that the latter fits the GRB HD with a reduced
$\chi^2_{\rm dof}\approx 2.14$, compared to $2.26$ for the standard
model. Nonetheless, these high values also show that the
use of GRBs for cosmological purposes is not yet mature enough
to carry out precision tests. However, in
attempting to assess which of these two models is
favored by the GRB data, we have found that several well-studied
criteria developed for this purpose all consistently point to the
$R_{\rm h}=ct$ Universe as being more likely to be correct
than $\Lambda$CDM, with a likelihood of $\sim 85-96\%$
versus $\sim 4-15\%$.

Another significant result of our study is the remarkable overlap
of the two best-fit curves in Figure~7. This feature is reminiscent
of a similar result from our earlier study of Type Ia SNe, particularly
Figure~4 in Melia (2012a). We believe that this
is not a coincidence because several studies have now shown that
$\Lambda$CDM is apparently mimicking the expansion history
implied by $R_{\rm h}=ct$. The most detailed discussion on this
issue has appeared in Melia (2012c; 2013). In these papers,
we presented several arguments for why the optimization of the
free parameters in $\Lambda$CDM always seems to indicate
an overall expansion of the Universe equal to what it would
have been in $R_{\rm h}=ct$.

Our final comment concerns the implications of this work on
the use of Type Ia SNe to study the cosmological expansion
at $z\la 2$. There is no question now that any comparative
analysis between competing cosmologies must be carried
with the re-calibration of the data for each assumed
expansion scenario, particularly when using standard candles
that rely on integrated quantities, such as the luminosity distance.
The Type Ia supernova luminosity cannot be determined independently of the
assumed cosmology---it must be evaluated by optimizing 4 parameters
simultaneously with those in the adopted model. This renders the data
compliant to the underlying theory.

Given how much better $R_{\rm h}=ct$
accounts for the cosmological data, such as the angular correlation of the cosmic
microwave background (Melia 2012d) and the redshift evolution of $H(z)$
(Melia 2013), not to mention the GRB HD we have studied in this paper,
we believe it is necessary to produce a Type Ia supernova Hubble Diagram
properly calibrated for the $R_{\rm h}=ct$ cosmology. Only then will it
be possible to properly compare the best-fit $\Lambda$CDM model directly
with $R_{\rm h}=ct$ at $z\la 2$. The payoff from this effort should not be
underestimated. We would know for certain whether the Universe is truly
now accelerating, or whether it continues expanding at a constant rate,
as it apparently has been doing from the beginning.

\vskip-0.2in
\acknowledgments
We thank the anonymous referee for his/her very constructive suggestions.
We also thank Z. G. Dai,
E. W. Liang, T. Lu, S. Qi, F. Y. Wang, M. Xu, and B. Zhang for helpful discussions.
XFW acknowledges the National Basic Research Program (``973" Program) of China
under Grant Nos. 2009CB824800 and 2013CB834900, the One-Hundred-Talents Program
and the Youth Innovation Promotion Association of the Chinese Academy of Sciences,
and the Natural Science Foundation of Jiangsu Province.
FM is grateful to Amherst College for its support through a John Woodruff Simpson Lectureship,
and to Purple Mountain Observatory in Nanjing, China, for its hospitality while this work
was being carried out. This work was partially supported by grant 2012T1J0011 from The
Chinese Academy of Sciences Visiting Professorships for Senior International Scientists,
and grant GDJ20120491013 from the Chinese State Administration of Foreign Experts Affairs.

\end{document}